# Adiabatic orientation of rotating dipole molecules in an external field

J. Bulthuis[1], J.A. Becker[2], R. Moro[3], V.V. Kresin[4]

[1] Department of Physical Chemistry and Laser Centre, Vrije Universiteit, De Boelelaan 1083, 1081 HV Amsterdam, The Netherlands

[2] Institut für Physikalische Chemie, Universität Hannover, D-30167 Hannover, Germany

[3] Department of Physical Sciences, Cameron University, Lawton, Oklahoma 73505, USA

[4] Department of Physics and Astronomy, University of Southern California, Los Angeles, California 90089-0484, USA

**Abstract**

The induced polarization of a beam of polar clusters or molecules passing through an electric or magnetic field region differs from the textbook Langevin-Debye susceptibility. This distinction, which is important for the interpretation of deflection and focusing experiments, arises because instead of acquiring thermal equilibrium in the field region, the beam ensemble typically enters the field adiabatically, i.e., with a previously fixed distribution of rotational states. We discuss the orientation of rigid symmetric-top systems with a body-fixed electric or magnetic dipole moment. The analytical expression for their "adiabatic-entry" orientation is elucidated and compared with exact numerical results for a range of parameters. The differences between the polarization of thermodynamic and "adiabatic-entry" ensembles, of prolate and oblate tops, and of symmetric-top and linear rotators are illustrated and identified.

## I. INTRODUCTION

The subject of orientation of dipolar molecules by static electric and magnetic fields has long occupied a prominent place in the study of molecular beams [1-3], and is currently enjoying renewed attention due to the introduction of "brute force" orientation methods [4,5] for stereospecific collision experiments [6,7] and spectroscopy [8], to the development of Stark deceleration techniques for the production and storage of slow molecules [9], to construction of electrostatic guides for cold molecules [10], and to electric and magnetic beam deflection experiments on polar clusters (see, e.g., [11-16] and references therein). A basic question can be stated as follows: for a beam of polar molecules passing through a region of applied external field, what is the resulting polarization, i.e., the ensemble average projection of the molecular dipole on the field axis? In the following, we will use the language of electric fields and dipole moments for conciseness, but the discussion is equally applicable to magnetic field effects on a cluster or molecule with a body-fixed magnetic dipole moment.

Theoretical treatments of the dielectric response of non-interacting polar gas molecules has a long history. For an ensemble of molecules in a gas and in the presence of an electric field $\mathcal{E}\hat{z}$, it was shown by Langevin [17] that the contribution of the permanent dipole $\mu_0$ to the polarization can be written as

$$p_z = \mu_0 \left[ \coth x - 1/x \right] \equiv \mu_0 \mathcal{L}(x), \tag{1}$$

where

$$x \equiv \mu_0 \mathcal{E} / k_B T. \tag{2}$$

$\mathcal{L}(x)$ is appropriately called the Langevin function. If $x \ll 1$, the polarization can be approximated by $p_z = \mu_0^2 \mathcal{E} / 3 k_B T$, generally referred to as the Langevin-Debye law. The Langevin-Debye law can also be applied to condensed phases, with the proviso that the field strength is then that of the self-consistent (e.g., cavity) field rather than of the external field. In the early stage of quantum theory that the Langevin-Debye law is still valid to a good approximation, if the quantization of the rotor energies is taken into account, as described, e.g., in the classic books [18-20] which also provide the original references.

The Langevin-Debye law is derived for a molecular ensemble thermalized inside a field. The circumstances are not the same for a transiting beam of polar molecules, that is, for the situation typically encountered in molecular beam experiments. In this case, the molecules have acquired their initial state distribution in an earlier zero-field region (e.g., the rotational temperature $T_{rot}$ produced at the beam source), and respond individually to adiabatic entry into the field. (Adiabaticity corresponds to many experimental situations, and will be assumed throughout). Although the fact that low-field response should still scale with the ratio $\mu_0 \mathcal{E}/T$ is expected on dimensional grounds [14], the numerical coefficient is no longer universal.

Restricting the present subject to rigid tops, the problem is to consider a polar molecule or cluster in a given rotational state $|n\rangle$ which has adiabatically entered the field region, and to evaluate its time-averaged orientation cosine, i.e., the expectation value of the angle between the molecular dipole axis and the field direction: $P_1|_n \equiv \langle \cos\theta \rangle_n = \langle \mu_z \rangle_n / \mu_0$. This quantity is then to be averaged over the initial state distribution, yielding the net orientation of the beam ensemble:

$$\overline{P_1} \equiv \overline{\langle \cos\theta \rangle} = \overline{\langle \mu_z \rangle} / \mu_0 \qquad (3)$$

In many situations one may assume a statistical Boltzmann distribution $\exp(-E_n / k_B T_{rot})$. In the end, the orientation can be used to compute the energy shift of the particle in the field, the deflecting or retarding force, etc.

The aforementioned Langevin-Debye expression corresponds to

$$\overline{P_1}^{L-D} = \frac{1}{3}\frac{\mu_0 \mathcal{E}}{k_B T}, \qquad (4)$$

but orientation under the adiabatic entry condition will always be *less*. The reason is that in the case of molecules in thermal equilibrium in the field, the ones with the dipole oriented along the field will have a lower energy and thus a higher statistical weight; this bias is absent for molecules passing through the field adiabatically.

For the case of a totally symmetric spherical top ($I_a=I_b=I_c$, or $A=B=C$ [21]) Bertsch *et al.* have shown both quantum-mechanically [22] and classically [23] that for weak fields,

$$\overline{P_1}^{sph} = \frac{2}{9}\frac{\mu_0 \mathcal{E}}{k_B T_{rot}}. \qquad (5)$$

This result was later generalized by Schnell *et al.* [14] to the case of an axially symmetric rotor (see below).

Since the theory is used to relate various microscopic and laboratory quantities (force experienced by the particles, field strength, dipole moments, rotational temperature of the beam), it is important to verify the accuracy of these expressions for a range of parameter values. This is the subject of Sec. II, which supplies a correction and some details for the derivation in Ref. [14], and pinpoints the source of the difference between the "adiabatic-entry" and "in-field equilibrium" orientations of symmetric-top dipoles in an external field. Furthermore, it compares the analytical results with a direct calculation of rotational Stark shifts for symmetric tops.

In contrast to the symmetric top case, to second order the orientation of a linear rotator in a weak external field originates exclusively from the lowest rotational quantum level. As a consequence, its adiabatic-entry orientation follows the Langevin-Debye law (with some quantum corrections). This is discussed in Sec. III, which also compares the Langevin form with a direct numerical calculation and with the strong-field limit analyzed in Ref. [24]. Section IV contains a summary and conclusions.

## II. SYMMETRIC TOP
### A. Dipole orientation of the adiabatic ensemble

The quantum numbers of rotational motion are $J,K,M$, where $J$ is the total rotational angular momentum quantum number, $K$ is the projection of $\mathbf{J}$ on the symmetry axis of the rotor, and $M$ is the projection of $\mathbf{J}$ along the field direction. The electric dipole is along the symmetry axis of the system. The external field induces precession of the top [14,23,25,26]. From second order perturbation theory [27], one finds that the orientation for a symmetric top is [28]

$$P_1\big|_{JKM} = \frac{MK}{J(J+1)} + \frac{\left[(J+1)^2 - M^2\right]\left[(J+1)^2 - K^2\right]}{(J+1)^3 (2J+1)(J+3)} \omega - \frac{(J^2 - M^2)(J^2 - K^2)}{J^3 (2J-1)(2J+1)} \omega, \qquad (6)$$

where

$$\omega \equiv \mu_0 \mathcal{E} / B . \qquad (7)$$

For $J = 0$, $P_1\big|_{J=0} = \tfrac{1}{3}\omega$. Averaging Eq. (6) over $M=-J,...J$, one finds [14,28]

$$P_1\big|_{\langle J \neq 0, K \rangle} = \frac{K^2}{3J^2(J+1)^2} \omega . \qquad (8)$$

As emphasized above, the average orientation is calculated by using the ensemble distribution which the particles had *before* entering the field region, hence by employing Boltzmann factors which contain the zero-field energies $E/k_B T_{rot} = Y^{-1}\left[J(J+1) - \kappa K^2\right]$. Here, following the usage of Ref. [29]

$$Y \equiv k_B T_{rot} / B \tag{9}$$

and the parameter $\kappa$ is defined as

$$\kappa = (C - A)/C \tag{10}$$

for a prolate top (with $B=C<A$ and therefore $\kappa<0$), and

$$\kappa = (A - C)/A \tag{11}$$

for an oblate top (with $B=A>C$ and $0<\kappa<0.5$, i.e., $C<A<2C$ with the upper limit representing a "flat ring" geometry). For a spherical top $\kappa=0$. Note that the consistent definition of $\kappa$ given here results in its coefficients having the opposite sign to those written down in Ref. [14].

The weighted average of the expression (8) then yields the following net orientation of the beam ensemble:

$$\overline{P}_1^{sym} = \frac{1}{3} \frac{\mu_0 \mathcal{E}}{B} \frac{1}{Z} \sum_{J=0}^{\infty} \sum_{K=-J}^{J} (2J+1) \frac{K^2}{J^2(J+1)^2} e^{-Y^{-1}\left[J(J+1)-\kappa K^2\right]} \tag{12}$$

with the partition function given by

$$Z = \sum_{J=0}^{\infty} \sum_{K=-J}^{J} (2J+1) e^{-Y^{-1}\left[J(J+1)-\kappa K^2\right]} . \tag{13}$$

[For the $J=K=0$ term one needs to set $K^2/J^2=1$ in Eq. (12), as follows from Eq. (6).]

Eq. (12) is equivalent to Eq. (58) of Ref. [14], where it was derived both classically and quantum-mechanically. Two adjustments compared to this reference should be noted: there the factors $(2J+1)$ were accidentally omitted, and the denominator $J^2(J+1)^2$ was replaced by $J^4$. It turns out that the latter replacement does not change the integrated result, Eqs. (14)-(16) below, but can lead to noticeable deviations when the sums (12),(13) are evaluated numerically.

If the temperature is sufficiently high and the rotational constant $B$ is not too large, sufficient $J$ states are populated to justify replacement of the summation over states in (12) and (13) by integration. In the angular momentum vector model, $K = \sqrt{J(J+1)} \cos\phi$, where $\phi$ is the angle between $\mathbf{J}$ and the dipole axis. The summation over $K$ can thus be replaced by integration over $\phi$ from 0 to $2\pi$.

Writing

$$\bar{P}_1^{sym} = \frac{\mu_0 \mathcal{E}}{k_B T_{rot}} z(\kappa) \tag{14}$$

and carrying out the integrations over $J$ and $\phi$, one finds:

$$z(\kappa)\big|_{\kappa>0} = \frac{1}{3\kappa}\left(-1+\kappa+\sqrt{\frac{1-\kappa}{\kappa}}\arcsin\left(\sqrt{\kappa}\right)\right), \tag{15}$$

$$z(\kappa)\big|_{\kappa<0} = \frac{1}{3\kappa}\left(-1+\kappa+\sqrt{\frac{1-\kappa}{-\kappa}}\operatorname{arcsinh}\left(\sqrt{-\kappa}\right)\right). \tag{16}$$

This result is identical to Eq. (60) of Ref. [14] if the sign of $\kappa$ in this reference is inverted. Note that to leading order the result is independent of the magnitude of the rotational constants and depends only on their ratio.

Expanding Eqs. (15) and (16) around $\kappa=0$ gives

$$\bar{P}_1^{sym} = \frac{2}{9}\frac{\mu_0 \mathcal{E}}{k_B T_{rot}}\left(1-\tfrac{1}{5}\kappa+...\right). \tag{17}$$

It is seen that the adiabatic-entry orientation for prolate tops ($\kappa<0$) is higher than for oblate tops ($\kappa>0$). The explanation of this result can be found in Eqs. (12) and (13). On one hand, since each term in the double sum in Eq. (12) contributes a positive amount to the orientation, this sum is smaller for prolate tops, because of the smaller $\exp(\kappa K^2/Y)$ Boltzmann factor. On the other hand, the partition function $Z$ in the denominator is also smaller for prolate tops. Since the effect of the Boltzmann factor in the numerator is suppressed by the $K^2/J^2(J+1)^2 < 1$ term which is absent in the partition function, the decrease in $Z$ "wins out" and enhances $\bar{P}_1^{sym}$ for the prolate case.

## B. Dipole orientation of the in-field equilibrium ensemble

The result of Eq. (17) is in sharp contrast with the Langevin-Debye formula, which has a factor of 1/3 instead of 2/9, and which, moreover, is independent of the asymmetry, i.e., of the parameter $\kappa$. The reason for these differences is to be found in the first term on the right-hand side of Eq. (6). It does not average to zero anymore if the ensemble is in thermal equilibrium within the field, because in this case molecules in states $|JKM\rangle$ with opposite signs of $M \cdot K$ have different statistical weights.

Although the standard result is well known, it is instructive to trace how it appears in the present situation. The ensemble-averaged orientation, with the first-order Stark shift $\Delta E^{(1)}$ now present in the Boltzmann factor, can be written as

$$\left(\bar{P}_1^{sym}\right)_{field} = \frac{1}{Z_{tot}} \sum_{J=0}^{\infty} \sum_{M=-J}^{J} \sum_{K=-J}^{J} P_1\big|_{JKM} \exp\left\{-Y^{-1}\left[J(J+1)-\kappa K^2\right] - \frac{\Delta E^{(1)}}{k_B T}\right\} \tag{18}$$

The shift comes from the first term in Eq. (6):

$$\Delta E^{(1)} = -\frac{MK}{J(J+1)} \mu_o \mathcal{E}, \tag{19}$$

while the second and third terms of $P_1\big|_{JKM}$ would contribute in the next order of the field strength $\mathcal{E}$. Expanding the exponential

$$\left(\bar{P}_1^{sym}\right)_{field} = \frac{1}{Z_{tot}} \sum_{J=0}^{\infty} \sum_{M=-J}^{J} \sum_{K=-J}^{J} P_1\big|_{JKM} \left(1 - \frac{\Delta E^{(1)}}{k_B T} + ...\right) e^{-Y^{-1}\left[J(J+1)-\kappa K^2\right]}, \tag{20}$$

and taking the sums over $M$, we obtain

$$\left(\bar{P}_1^{sym}\right)_{field} = \frac{1}{3} \frac{\mu_0 \mathcal{E}}{Z} \sum_{J=0}^{\infty} \sum_{K=-J}^{J} \left[\frac{1}{k_B T} \frac{K^2(2J+1)}{J(J+1)} + \frac{1}{B} \frac{K^2(2J+1)}{J^2(J+1)^2}\right] e^{-Y^{-1}\left[J(J+1)-\kappa K^2\right]} \tag{21}$$

To the same order in $\mathcal{E}$, $Z_{tot}$ may be replaced by $Z$. Analogously to Eq. (12), for $J=K=0$ the first term in square brackets vanishes, and in the second term $K^2/J^2$ is set equal to 1.

Note that in the adiabatic-entry ensemble, Eq. (12), only the second term in the square brackets was present. Again replacing the summations by integrations and making the same substitution of variables as in the evaluation of Eq. (12), one discovers that the additional term in Eq. (21) precisely cancels out the $\kappa$ dependence of the orientation. As a result, only the factor 1/3 survives, meaning that the Langevin-Debye equation is recovered. This is of course not surprising, since the derivation outlined above is a slightly modified version of Kronig's original treatment [30].

The sizable increase from the adiabatic-entry value of the orientation, $\bar{P}_1^{sym}$ [Eqs. (14)-(17)] to the Langevin-Debye value $\left(\bar{P}_1^{sym}\right)_{field} = \mu_0 \mathcal{E} / 3 k_B T_{rot}$ is, as stated above, due to the contribution of the first term in brackets in the sum in Eq. (21). Its role in making the end result independent of $\kappa$ can be seen qualitatively from the fact that it contributes more for oblate than for prolate tops [the ratio $K^2/J(J+1) \sim 1$ does not strongly suppress the effect of the $\exp(\kappa K^2/Y)$ Boltzmann factor relative to the partition function in the denominator] just canceling the opposite behavior of the second term discussed at the end of Sec. IIA.

### C. Comparison with exact evaluation

The calculations given above can now be compared with a virtually exact diagonalization of a truncated Hamiltonian matrix for a rigid rotor in an electric field. The truncation is taken such that a converged result is obtained, i.e., extension of the matrix has no effect on the energies of the states of interest, here chosen as those with statistical weights exceeding $10^{-6}$. This determines $J_{max}$, the maximum $J$ value for states with a given $K$ and $M$. In practice, extension of the matrix by states with $J$ exceeding $J_{max}$ by 10 is found to be adequate to get a converged result. In this way the accurate energies and eigenvectors for all parent states with a given $K$ and $M$ can be calculated by a single diagonalization. This is much more efficient than splitting up the matrix into smaller matrices for each parent $J$; the approach used in Refs. [28,36]. The procedure of computing the average orientation from the eigenvectors is described is described in Ref. [28].

For the case of adiabatic entry, Fig. 1 shows that for the chosen parameters the average orientation of a symmetric top as a function of $C/A$ for an oblate top and $A/C$ for a prolate top, is nearly independent of the degree of approximation. The approximate calculations are based on a numerical evaluation of Eq. (12) for $J$ up to 200, and on Eq. (14). For $B$=0.1 cm$^{-1}$, the exact curves deviate only slightly from those in Fig. 1.

The numerical summation of Eq. (12) always gives a slightly lower orientation than Eq. (14), i.e., closer to the exact result, because the additional approximation of replacing the summation by an integration is not made. The deviations of both approximations from the exact diagonalization result increases with $A/C$ and is also roughly proportional to $B$.

As $B$ gets larger, the orientation from the exact calculation decreases slightly, but deviation from the approximate calculations, which are independent of $B$, remains small as long as $B$ is not increased to uncommonly large values (>>1 cm$^{-1}$). For increasing $x = \mu_0 \mathcal{E}/k_B T_{rot}$ deviations become progressively larger, as shown in Fig. 2.

In agreement with Fig. 1, exact calculations on methyliodide [4,31,32], with $A/C$>10, showed that the average orientation was just below the Langevin-Debye limit (4).

At the top of Fig. 1 the Langevin-Debye limit, Eq. (2), is also shown and compared with an exact numerical diagonalization and with the sum in Eq. (21). For the chosen parameters, the deviations are very small. This confirms that for an ensemble thermalized within the field, the dependence of the polarization on the rotation parameters is essentially absent. As $x$ increases, the Langevin-Debye value of the orientation remains within a few percent of the exactly computed one, while the approximate sum in Eq. (21) again increasingly deviates from the latter (it can even yield orientations that exceed the Langevin-Debye limit, which is indicative of the inadequacy of keeping only second-order perturbation terms with increasing field strengths and/or decreasing temperatures).

## III. LINEAR ROTATOR

The response of linear dipoles to an external field was analyzed in the very early days of quantum mechanics. It is instructive, though, to look at this case vis-à-vis that of the symmetric top considered above. For a linear molecule, the first order Stark energy is zero. As a consequence, in this special case the low-field susceptibility is the same in the adiabatic-entry and in-field equilibrium situations.

Quantum mechanically, in second-order perturbation theory the orientation for a state $|J,M\rangle$ is given by Eq. (4) with $K=0$. Averaging over all $M$ values results in zero orientation for all $J \neq 0$ states and a net orientation of $\omega/3$ for $J=0$ [this also follows from setting $K=0$ in Eq. (8)]. The ensemble average is therefore simply proportional to the population of the ground rotational state: $\bar{P}_1^{lin} = Z^{-1}(\omega/3)$, with $Z = \sum_{J=0}^{\infty}(2J+1)\exp[-J(J+1)/Y]$. Approximating the sum over states with the help of the Euler-Maclaurin summation formula, we obtain the long-established result [18]

$$\bar{P}_1^{lin} = \frac{\mu_0 \mathcal{E}}{3k_B T_{rot}}\left[1 - \frac{B}{3k_B T_{rot}} + \frac{1}{36}\left(\frac{B}{k_B T_{rot}}\right)^2 + ...\right]. \tag{22}$$

Eq. (22) is identical to the classical Langevin-Debye equation, apart from the correction terms in brackets. For small $B$ values it lies only slightly higher than the result of an exact numerical calculation, as shown in Fig. 3. The decrease in orientation with increasing $B$ is connected with the fact that it is dependent on the distance of the $J=0$ level to the $J\geq 1$ levels, and for increasing $B$ the coupling to these higher levels decreases.

Recall that at sufficiently high temperatures the orientational polarization of symmetric tops discussed in Sec. II is controlled by high *J* states and hence can be approximated by the classical (or semiclassical) dynamics of a dipole rotator [14,23,26]. In contrast, the classical limit of the polarization of a rotating linear dipole is [33] $p_z \propto \mu_0^2 \mathcal{E} (3\cos^2 \varphi - 1)$, where $\varphi$ is the angle between the rotation axis and the field; upon averaging $p_z$ over all angles one finds zero net polarization. The qualitative difference between the two classes of rotors is that the symmetric one is gyroscopic, and the linear one is not. This was already noted by Pauli in 1921 who pointed out that classically only nonrotating linear molecules, "which execute small vibrations about a position of rest" [20,34], can contribute to net polarization.

The validity of Eq. (14) is subject to the condition that $\mu_0 E/k_B T \ll 1$. For higher fields, the adiabatic-entry and in-field-equilibrium curves again diverge, with the orientation in the former case always lower, as expected on the basis of arguments given above for symmetric top molecules. This is illustrated in Fig. 4.

In the in-field equilibrium case, the orientation of *all* rotors should follow the Langevin function $\mathcal{L}(x)$ as long as the rotational energy spacing is much less than $k_B T$ [35]. This is borne out by an exact computation for a linear dipole, as is also illustrated in Fig. 4: the deviation becomes non-negligible only for low values of *Y*, i.e., for $k_B T \sim B$.

As for adiabatic-entry polarization, one can take advantage of the fact that at high field strength and low rotational temperature, an increasing number of states are converted to pendular states. Recently Friedrich [24] employed the correlation between the free-rotor states and the harmonic librator limit to derive an analytic approximation for the orientation and alignment of linear molecules. His result (Eq. (12) of Ref. [24]) coincides with the exact calculation for low values of *Y* or high values of *ω*. In this way, Eq. (22) and Friedrich's result bracket the adiabatic-entry response at the opposite ends, see Fig. 4. It would certainly be desirable to have an analytical expression for the intermediate range of *Y* and *ω* values, where neither approximation is applicable. This would require the use of higher than second-order perturbation theory; however, only perturbations of even order, leading to contributions of odd order in ω, enter in the orientation of a linear dipole, and thus at the minimum 4[th] order perturbation theory would be required. Thus (at least for the moment) it remains more practical to carry out exact numerical calculations for the intermediate case.

## IV. CONCLUSIONS

In studies of oriented dipolar molecules and clusters, obtained by applying an electric or magnetic field to a molecular beam, no thermal equilibrium exists while the particles pass the field region. Instead, commonly the field is entered adiabatically, so the populations of the rotational levels which are relevant for determining the average orientation remain the same as they were in the zero-field region.

As a consequence, the ensemble-averaged orientation for adiabatic entry is always lower than that for an ensemble in equilibrium in the field. This is easily seen by realizing that the populations of states with a negative Stark effect (which give a positive contribution to the orientation) increase if the ensemble is in thermal equilibrium in the field. Specifically, the orientation of spherically symmetric tops is still proportional to $x(=\mu_0\mathcal{E}/k_B T_{rot})$ for small values of $x$, but the proportionality factor is not $\frac{1}{3}$, as in the textbook Langevin-Debye limit, but is predicted to be $\frac{2}{9}$ [22,23].

For the more general case of a symmetric top, an approximate analytical expression for $\bar{P_1}$ based on second-order perturbation theory demonstrates that the adiabatic-entry orientation for oblate tops is even smaller than for the spherical top, while for prolate tops the orientation increases with the ratio of rotational constants *A/C* (with *B=C*), and for large *A/C* approaches the Langevin-Debye value [14]. This is in contrast to an ensemble thermalized within the field, where there is no dependence on the *A/C* ratio. The origin of the difference between oblate and prolate tops can be identified qualitatively from the behavior of the relevant statistical sums.

We have elucidated the derivation of the aforementioned analytical approximation and explored its applicability by comparing the results with those of exact numerical calculations for a range of conditions. The conclusion is that for commonly observed values of rotational constants, the approximations are very good for $x$ values of up to ~0.3.

Many symmetric top molecules that have been investigated in molecular beams are prolate tops with a rather large *A/C* ratio, and this may be a reason why the fundamental difference between the average orientation of adiabatic-entry and thermal-equilibrium ensembles was not accentuated for a long time. This difference, and the marked variation of $\bar{P_1}$ as a function of shape, must be kept in mind for the interpretation of deflection experiments on polar and magnetic nanoclusters, because clusters can exhibit a diversity of oblate and prolate shapes.

For a linear rotor, both types of ensemble produce the same orientation for $x\ll1$, but for increasing field strengths or low temperatures the differences become increasingly larger, with the adiabatic-entry orientation again falling below the field-thermalized value. For the latter case, the classical Langevin function gives a very good description down to quite low temperatures, but for the adiabatic-entry orientation no generally applicable function is known. Here $\overline{P}_1$ is bracketed between the Langevin function for low fields/high temperatures and the function derived in [24] for strong fields/low temperatures (based on the harmonic librator limit).

No analytical solution for adiabatic-entry orientation for asymmetric tops is presently available. Because of the additional *K*-mixing, a second-order perturbation approach is intractable, and one has to resort to numerical diagonalization [36,37] or molecular dynamics simulations [38]. There are suggestions that because of the nascent chaotic character of asymmetric top motion [39-41] and/or because of the general action of avoided crossings in the coupled Stark (Zeeman)-rotational level diagram [12], the statistical Langevin-Debye behavior will be restored. For several asymmetric tops the adiabatic-entry orientation was calculated exactly and found to be close to the Langevin-Debye equation [36,37]. Curiously, this is the case even for the water molecule (where the Stark-split rotational energy levels are well-ordered and well-separated) already at relatively low rotational temperatures [42].

## ACKNOWLEDGMENTS

J.B. thanks Dr. G. van der Zwan for helpful discussions on statistical thermodynamics issues. The work of V.K. was supported by the U.S. National Science Foundation under Grant No. 0652534.

# Figures

**Fig. 1.** Orientation of the symmetric top ($\mathcal{E}$=50 kV/cm, $\mu_0$=2 D, $T_{rot}$=50 K and $B$=1 cm$^{-1}$; $x \approx 0.05$), as a function of the C/A (oblate) and A/C (prolate) ratio, for an ensemble in thermal equilibrium in the electric field (top set of lines), and for an ensemble for which the molecules enter the field adiabatically (lower set of curves).

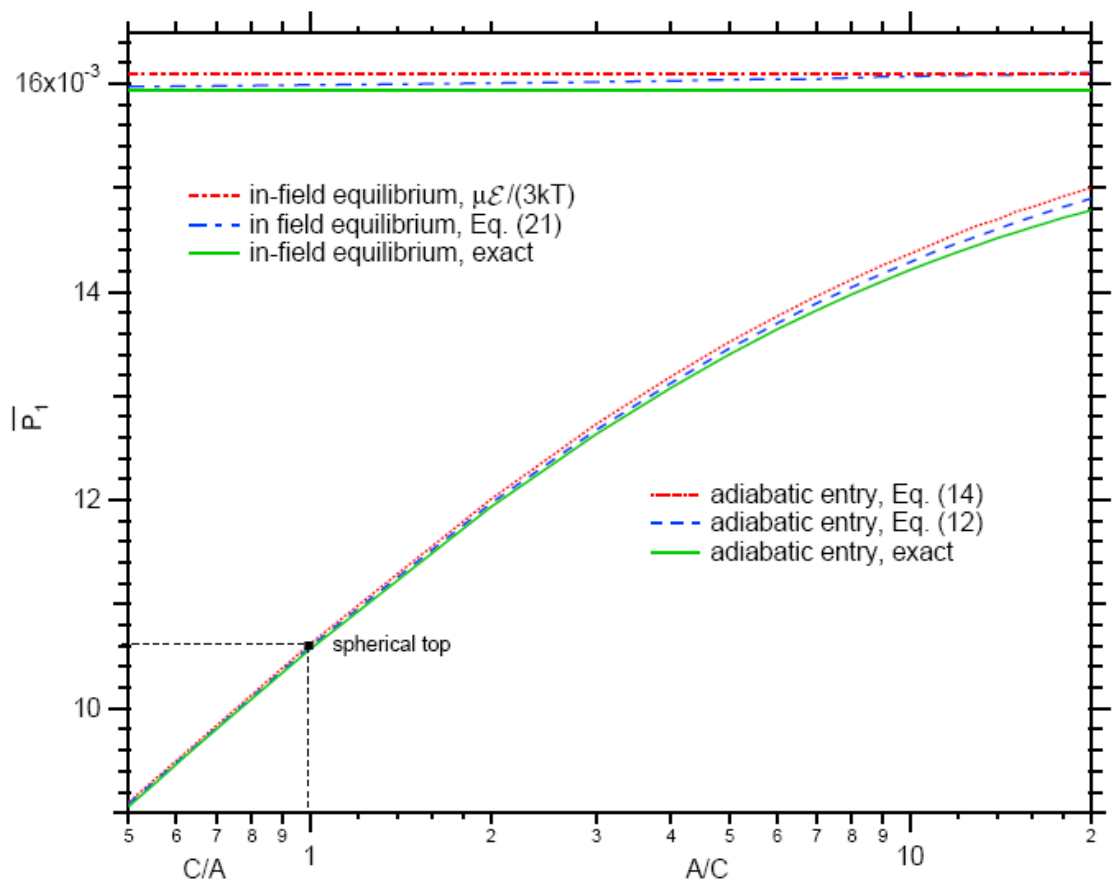

**Fig. 2.** Adiabatic-entry orientation of symmetric top molecules ($\mathcal{E}$=50 kV/cm, $\mu_0$=2 D, B=0.1 cm$^{-1}$; $\omega$=50) calculated using Eqs. (12) and (14), divided by the orientation from an exact calculation, for four different temperatures, corresponding to $x$ values of 0.14, 0.25, 0.71, and 1.42 (from high to low temperature). Of the pairs of curves, the lower curve corresponds to direct numerical summation of Eq. (12).

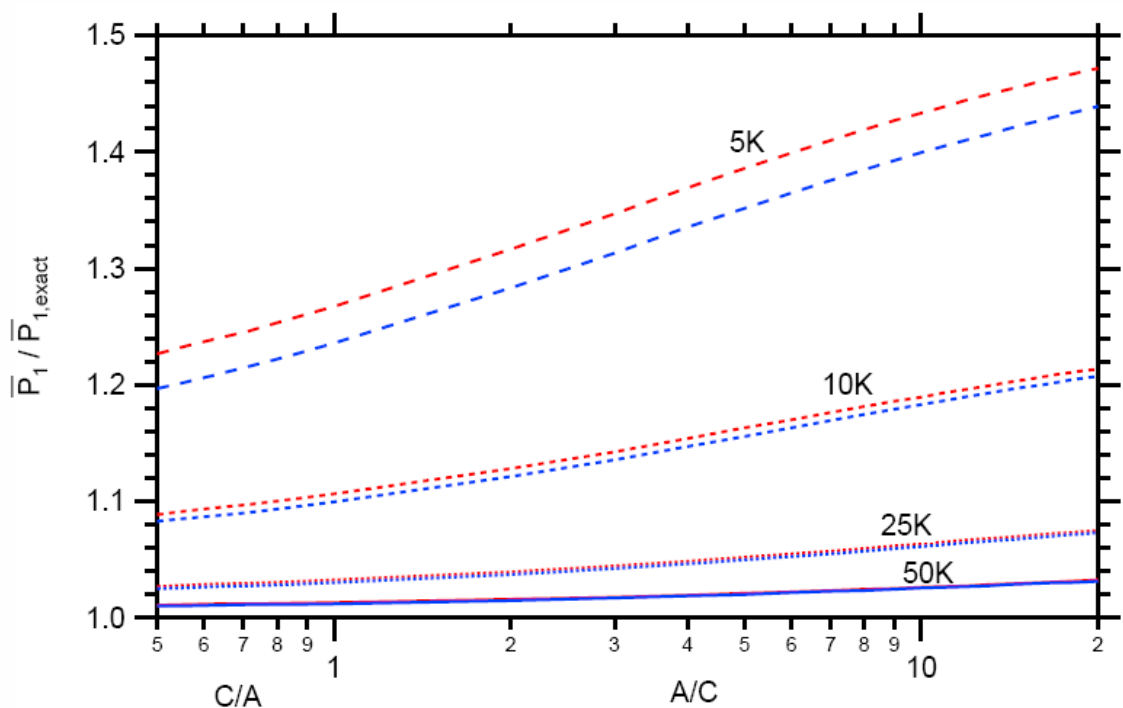

**Fig. 3** Orientation of linear dipole molecules ($\mathcal{E}$=50 kV/cm, $\mu_0$=2 D, $T_{rot}$=50 K) as a function of the rotation constant $B$ for an in-field equilibrium ensemble and for adiabatic entry. The curve from the exact calculation for the in-field equilibrium case virtually coincides with the curve from Eq. (22). Note that to second-order accuracy, for a linear rotor there is no distinction between in-field equilibrium and adiabatic entry. The $B$-dependent correction in Eq. (22) becomes significant for rotational constants above a few wavenumbers (for example, for HCl, $B\approx$10 cm$^{-1}$, the correction at T=50 K is nearly 10%).

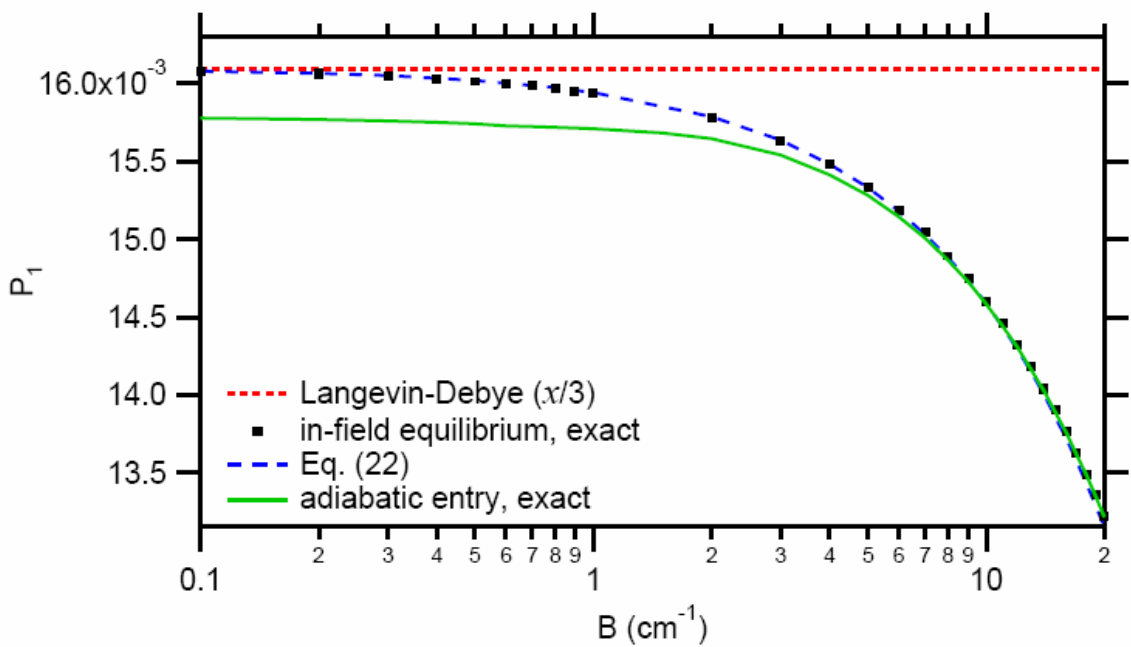

**Fig. 4.** Orientation of a linear dipole ($\mu_0$=2 D, $B$=0.1 cm$^{-1}$) for three different temperatures, 0.3 K, 2.9 K, and 14.4 K, corresponding to $Y = k_B T_{rot} / B$ =2, 20, and 100, respectively, as a function of the reduced electric field parameter $\omega = \mu_0 \mathcal{E} / B$. For each of the temperatures, five curves are shown: the Langevin equation; its Langevin-Debye limit; the exact result for an ensemble in thermal equilibrium within the field; the exact result for the adiabatic entry ensemble; and the equation from Ref. [24] based on the harmonic-librator limit for the adiabatic-entry ensemble. For $Y$=100, the curves of the exact in-field equilibrium result and the Langevin function coincide.

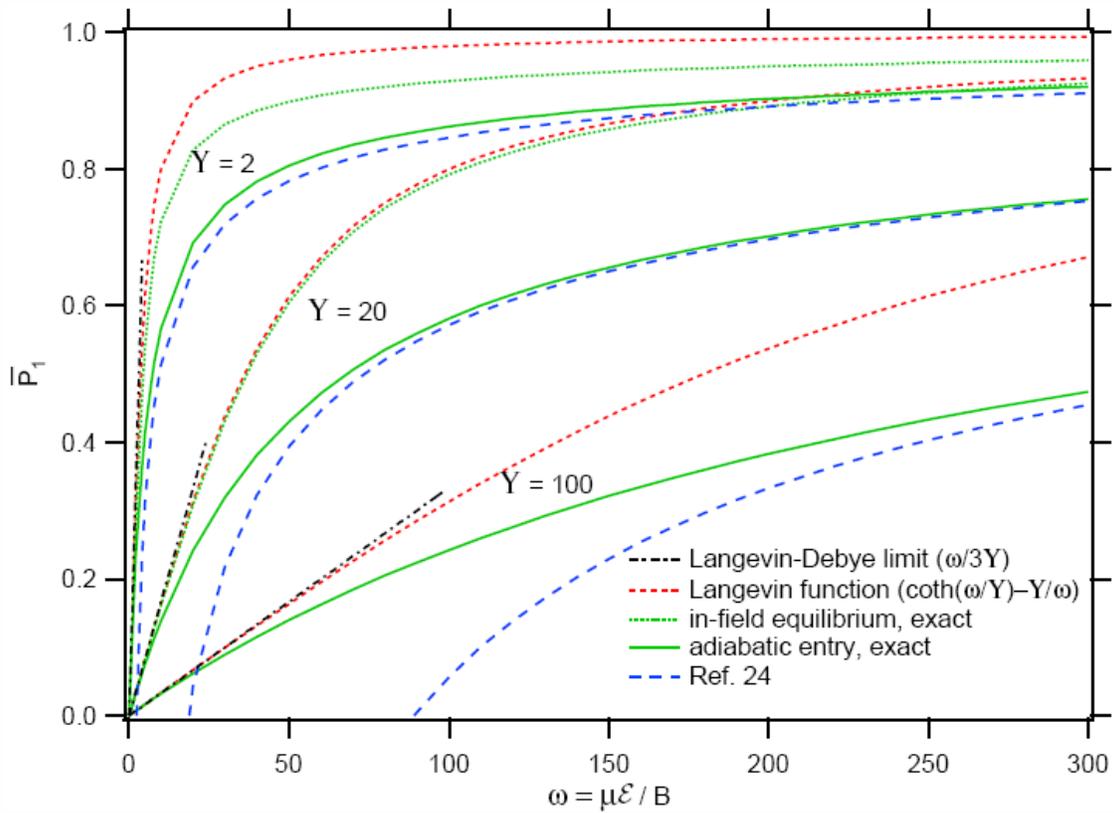